\documentstyle[12pt]{article}
\def\be{\begin{equation}}
\def\ee{\end{equation}}
\def\bea{\begin{eqnarray}}
\def\eea{\end{eqnarray}}
\begin{document}
\begin{titlepage}
\vglue -1cm
\begin{flushright}
ISN 96--94\\ nucl-th/9608058
\end{flushright}
\vfil\vfil
\centerline{\large\bf MULTIQUARK SYSTEMS}\vskip .1cm
\centerline{\large\bf  IN A CONSTITUENT QUARK MODEL}\vskip .1cm
\centerline{\large\bf  WITH CHIRAL DYNAMICS\footnote{Contribution to
the Workshop ``Quark Confinement and the Hadron Spectrum II'', Como,
Italy, June 26--29, 1996, to appear in the Proceedings, ed. Nora
Brambilla, World Scientific.} }
\vskip 1cm
\centerline{
{\bf S.~Pepin}\footnote{e-mail: {\tt u2162et@vm1.ulg.ac.be}}   
and {\bf Fl.~Stancu}\footnote{e-mail: {\tt u216207@vm1.ulg.ac.be}}
}  
\centerline{\small Universit\'e de Liege, Institut de Physique  B.5,} 
\centerline{\small Sart Tilman, B--4000 Li\`ege 1, Belgium}
\vskip .1cm 
\centerline{and}
\vskip .1cm
\centerline{ {\bf M. Genovese}\footnote{e-mail: {\tt
genovese@isnmv1.in2p3.fr}}
\footnote{Supported by the EU Program ERBFMBICT950427}  and {\bf
J.-M.Richard}\footnote{e-mail: {\tt jmrichar@isnnx2.in2p3.fr}} }
\centerline{\small Institut des Sciences Nucl\'eaires}
\centerline{\small Universit\'e Joseph Fourier--IN2P3-CNRS}
\centerline{\small 53, avenue des Martyrs, F-38026 Grenoble Cedex,
France}
\vfil
\begin{abstract}
{\noindent We discuss the stability of multiquark systems within the
recent model of Glozman {\sl et al.\/} where the chromomagnetic
hyperfine interaction is replaced by pseudoscalar-meson exchange. We
find that such an interaction binds a heavy tetraquark system $QQ\bar
q\bar q$ ($Q=c,\,b$ and $q=u,\,d)$ by
$0.2-0.4\;$GeV. This is at variance with results of previous models
where
$cc\bar q\bar q$ is unstable.}
\end{abstract}
\vfil
\end{titlepage}

The baryon spectrum has recently been revisited by using chiral models
which include meson-exchange forces between quarks
\cite{Glozman96,Dziembowski96,Valcarce96}. A generic Hamiltonian
summarising the above references reads
\bea\label{defH}
 H=&&\!\!\!\sum_i{\vec{\rm p}_i^{\,2}\over 2m_i}-{3\over16}
    \sum_{i<j} 
\tilde{\lambda}_i^{\rm c}\!\cdot\!\tilde{\lambda}_j^{\rm c}\,
               V_{\rm conf}(r_{ij})\nonumber\\
&&{}-\sum_{i<j}
\tilde{\lambda}_i^{\rm c}\!\cdot\!\tilde{\lambda}_j^{\rm c}\,
\vec{\sigma}_i\!\cdot\!\vec{\sigma}_j\,V_{\rm g}(r_{ij})+
\sum_{i<j}\tilde{\lambda}_i^{\rm F}\!\cdot\! 
\tilde{\lambda}_j^{\rm F}\,
\vec{\sigma}_i\!\cdot\!\vec{\sigma}_j \,V_{\rm F}(r_{ij}),
\eea
where $m_i$ is the constituent mass of the quark located at $\vec{\rm
r}_i$;
 $r_{ij}=\vert \vec{\rm r}_j-\vec{\rm r}_i\vert$ denotes the
interquark distance; $\vec{\sigma_i}$, $\tilde{\lambda}_i^{\rm c}$,
$\tilde{\lambda}_i^{\rm F}$ are the spin, colour and flavour
operators, respectively. Spin-orbit and tensor components may
supplement the above spin-spin forces for studying orbital
excitations.

The last term in $H$ represents the meson exchange. An implicit sum
over F is understood, where ${\rm F}=1,\,2$ and 3 corresponds to
$\pi$, 
${\rm F}=4,\,5,\,6$ and 7 to $K$, ${\rm F}=8$ to $\eta$, and  ${\rm
F}=0$ to
$\eta'$. When $V_{\rm F}=0$, we have a standard constituent quark
model. 

The confining term $V_{\rm conf}$ usually consists of a Coulomb plus
a linear term,
\be\label{Vconf} V_{\rm conf}=-{a\over r}+b r,
\ee and is sometimes approximated by a harmonic potential, with
possible constant terms.

The third term in $H$ is often understood as the chromomagnetic
analogue of the Breit--Fermi term of atomic physics.  The radial
shape of  $V_{\rm g}$ is taken as being of very short range. For
mesons, $\tilde{\lambda}_1^{\rm c}\!\cdot\!
\tilde{\lambda}_2^{\rm c}=-16/3$, and a positive $V_{\rm g}$, as in
the one-gluon-exchange model, shifts each vector meson above its
pseudoscalar partner, for instance $D^*>D$ in the charm sector. For
baryons, where  $\tilde{\lambda}_1^{\rm c}\!\cdot\!
\tilde{\lambda}_2^{\rm c}=-8/3$ for each quark pair, such a positive
$V_{\rm g}$ pushes the spin 3/2 ground states up, and the spin 1/2
down, for instance
$\Delta >N$. In the simplest models, $V_{\rm
g}\propto\delta^{(3)}(\vec{\rm r})$ is treated in first order.
Adopting a finite-range parametrisation allows one to treat $V_{\rm
g}$ non-perturbatively when solving the Schr{\"o}dinger equation.

It has been long recognised that explicit fitting of light mesons and
baryons in potential and bag models requires a large strength for the
chromomagnetic term. A possible remedy is to introduce mesonic loops.
For instance in the work of Myhrer {\sl et al.\/} \cite{Myhrer81} or
Cottingham {\sl et al.\/} \cite{Cottingham81}, the $\Delta-N$
splitting is shared  between pion loops and chromomagnetism in about
equal parts. The complementarity between gluon-exchange and
pion-field effects arises naturally in models where the bag
containing the quarks is surrounded by a pion cloud. Such a model is
the ``little bag'' of Brown and Rho \cite{Brown79}, where the pion
field is strictly restricted to stay outside the bag. In the ``cloudy
bag'' of Thomas and collaborators \cite{Thomas84}, the pion field is
allowed to penetrate the bag. 

Non-relativistic versions of the pion-exchange effect are the models
introduced by Weber {\sl et al.\/} \cite{Weber86}, and by Glozman and
Riska
\cite{Glozman96a}.  In these models,  pions and other mesons are
directly exchanged between the quarks, and thus travel through the
very interior of the hadron.

The explicit  pion-exchange contribution to the last term in
Eq.~(\ref{defH}) reads
\be\label{pionexc}
\sum_{i<j}\vec{\sigma}_i\!\cdot\!
\vec{\sigma}_j\,\vec{\tau}_i\!\cdot\!\vec{\tau}_j
{g^2\over 4\pi}{1\over 4m^2}\left[\mu^2{\exp(-\mu r_{ij})\over
r_{ij}}-4\pi\delta^{(3)}(r_{ij})\right],
\ee
where $\mu$ is the pion mass. A coupling constant $g^2/4\pi=0.67$  at
the quark level corresponds to the usual strength $g_{\pi
N\!N}/4\pi\simeq14$ for the Yukawa tail of the nucleon--nucleon
($N\!N$) potential. When constructing $N\!N$ forces from meson
exchanges, one disregards the short-range term in
Eq.~(\ref{pionexc}), for it is hidden by the hard core, and anyhow
the potential in that region is parametrised empirically. Similarly,
when T{\"o}rnqvist \cite{Tornqvist91}, Manohar and Wise
\cite{Manohar93}, or Ericson and Karl \cite{Ericson93}  consider pion
exchange in multiquark states, they have in mind the Yukawa term
$\exp(-\mu r)/r$ acting between two well separated quark clusters.
For similar reasons Weber {\sl et al.\/}
\cite{Weber86} ignore the delta-term too. Therefore it is  somewhat
of a surprise to see the delta-term of Eq.~(\ref{pionexc}) taken
seriously, and with an {\sl ad-hoc\/} regularisation playing a
crucial role in the quark dynamics
\cite{Glozman96,Dziembowski96,Valcarce96}.

In the work of Glozman, Papp and Plessas \cite{Glozman96},  the
chromomagnetic term is entirely omitted ($V_{\rm g}=0$) and a weak
linear confinement is supplemented by $\pi$,
$\eta$ and  $\eta'$ exchanges. The model is used to estimate the
spectrum of $N$ and $\Delta$ baryons. The explicit form of $H$
integrated in the spin--flavour space is 
\be
H=H_0+{g^2\over48\pi m^2}
\left\{
\normalbaselineskip=20pt
\matrix{
&\!\!\!\!\!15V_\pi-V_\eta-2\left({g_0/ g}\right)^2
V_{\eta'}\quad\hbox{for}\quad N\cr 
&\!\!\!\!\!\phantom{1}3V_\pi+V_\eta+2
\left({g_0/g}\right)^2V_{\eta'}\quad\hbox{for}\quad\Delta\cr
}\right.
\ee
with
\bea
&&\!\!\!\!H_0=\sum_i m_i+\sum_i{\vec{\rm p}_i^{\,2}\over 2m_i}+
{C\over2}\sum_{i<j}r_{ij}\\
&&\!\!\!\!V_\mu=\Theta(r-r_0)\mu^2{\exp(-\mu r)\over r}-
{4\epsilon^3\over\sqrt\pi}\exp(-\epsilon^2(r-r_0)^2); 
\ (\mu=\pi,\;\eta,\;\eta').
\eea
The parameters are $m=0.337\;$GeV, $C/2=0.01839\;$GeV$^2$,
$g^2/4\pi=0.67$, $(g_0/g)^2=1.8$, $\epsilon=0.573\;$GeV, 
$r_0=2.18\;{\rm
GeV}^{-1}$, and
$\mu=0.139$,  $0.547$, $0.958\;$GeV for $\pi$, $\eta$ 
and $\eta'$, respectively.

When the meson-exchange terms are switched off, the $N$ and $\Delta$
ground states are degenerate at 1.63 GeV. When the coupling  is
introduced, the wave function is modified. We have performed crude
variational estimates with Gaussian wave functions, and  reproduced
the results of the more elaborate Faddeev calculation of
Ref.~\cite{Glozman96}. For the nucleon, we found that the
spin-independent part $H_0$ of the Hamiltonian  gives a contribution
of
$2.1\;$GeV, and receives a large
$-1.2\;$GeV correction from meson exchange.  For the $\Delta$ ground
state, the contribution of $H_0$ and meson exchange parts are
$1.9\;$GeV and $-0.6\;$GeV, respectively. Thus one ends up with a
reasonable value for the $\Delta-N$ splitting, close to $0.3\;$GeV.

But dramatic effects occur when the model is applied to  mesons  or
to multiquark systems containing heavy quarks. When the meson mass
$\mu$ reaches values of 2 or $3\;$GeV as for the $\eta_c$ or the
$D$, the two terms in Eq.~(\ref{pionexc}) basically cancels each
other. Moreover we expect little $c\bar c\leftrightarrow q\bar q$
mixing in the
$\eta_c$, so very weak coupling of $\eta_c$ to a light quark $q$.
Hence the most natural extension of the model (\ref{defH}) to a
combination of light and charmed quarks restrict meson exchange to
the former ones. Then:

{\sl 1.} The $D$ and $D^*$ mesons are degenerate. An average mass
$M(D)\simeq2\;$GeV can be obtained from $H_0$ if the charmed quark is
given a mass value of $m_c\simeq 1.35\;$GeV.

{\sl 2.} A reasonable splitting is obtained between the isoscalar
$\Lambda_c$ and the (degenerate) $\Sigma_c$ and $\Sigma_c^*$ baryons
of quark content $(cqq)$. The masses are $\Lambda_c= 2.32\;$GeV, and 
$\Sigma_c=\Sigma_c^*=2.48\;$GeV.

{\sl 3.} Another consequence is that the $(\bar c\bar c qq)$
multiquark is easily bound provided the light diquark is in a
spin--isospin  $S=0$, $I=0$  state. A crude trial wave-function (a
Gaussian for each internal Jacobi coordinate) is sufficient to give a
binding energy as large as
\be
2(\bar c q)-(\bar c \bar c qq)= 0.18\;\hbox{GeV}.
\ee
For $(\bar b\bar b q q)$ we obtain in the same way a binding energy
\be
2(\bar c q)-(\bar c \bar c qq)= 0.22\;\hbox{GeV},
\ee
which becomes about twice larger for more realistic potentials
including a Coulomb term in the central part, as per
Eq.~(\ref{Vconf}).

 A spin--isospin  $S=0$, $I=0$ state implies a colour
$\bar 3$ for the $qq$ diquark, and thus an $S=1$, $I=0$ and colour 3
state for 
$(\bar Q\bar Q)$, which thus takes advantage of the confining force.
It was  indeed  shown that in a flavour-independent potential (no
hyperfine interaction)
$(\bar Q\bar Q qq)$ becomes stable if the mass ratio $m(Q)/m(q)$ is
large enough
\cite{Ader82}; then, for realistic potentials, stability occurs more
likely for
$(\bar b\bar b qq)$ than for $(\bar c\bar c qq)$
\cite{Zouzou86,Silvestre93}. Note that the entire
$\bar Q\bar Q qq$ system discussed above has $S=1$ and $I=0$. When
the masses of
$\bar Q$ and $q$ are comparable, the dynamics mixes the triplet and
sextet states of diquarks \cite{Brink94}.

It was underlined by T{\"o}rnqvist \cite{Tornqvist91}, and Manohar
and Wise
\cite{Manohar93} that one-pion exchange might favour binding of
heavy-flavour configurations. These authors, however, proposed
quantum numbers $(S=0,\; I=1)$ or $(S=1,\; I=0)$ for the light
diquark, so that a negative
$\vec{\sigma}_1\!\cdot\!\vec{\sigma}_2\,
\vec{\tau}_1\!\cdot\!\vec{\tau}_2$ makes the Yukawa tail $\exp(-\mu
r)/r$ attractive. This implies a colour 6 for a $qq$ diquark with
relative angular momentum $\ell=0$, in order to fulfil the Pauli
principle. Thus $\bar Q\bar Q$ is in a colour $\bar 6$ state, at
variance with the considerations above.

{\sl 4.} Presently we are investigating whether other multiquark
systems are predicted to be stable in our simple extension of the
model of Glozman {\sl et al.\/} In particular we are studying the
$(ccqqqq)$ system. More details will be given in a forthcoming
publication~\cite{Pepin96}.
%
\section*{Acknowledgements} The support of the CNRS--FNRS cooperation
program is gratefully acknowledged.
%
\def\NCA{{\em Nuovo Cimento}}
\def\NIM{{\em Nucl. Instrum. Methods}}
\def\NIMA{{\em Nucl. Instrum. Methods} A}
\def\NPB{{\em Nucl. Phys.} B}
\def\NPA{{\em Nucl. Phys.} A}
\def\PLB{{\em Phys. Lett.}  B}
\def\PRL{{\em Phys. Rev. Lett.}}
\def\PRD{{\em Phys. Rev.} D}
\def\PRC{{\em Phys. Rev.} C}
\def\ZPC{{\em Z. Phys.} C}


\begin{thebibliography}{99}
%
\bibitem{Glozman96}{L.Ya. Glozman, Z. Papp and W. Plessas,
\PLB\ {\bf 381}, 311 (1996).}
%
\bibitem{Dziembowski96}{Z. Dziembowski, M. Fabre de la Ripelle 
and G.A. Miller, \PRC\ {\bf 53}, 2038 (1996).}
%
\bibitem{Valcarce96}{A. Valcarce, P. Gonz\'alez, 
F. Fern\'andez and V. Vento, 
\PLB\ {\bf 367}, 35 (1996).}
%
\bibitem{Myhrer81}{F. Myhrer, G.E. Brown and Z. Xu, 
\NPA\ {\bf 362}, 317 (1981).}
%
\bibitem{Cottingham81}{W.N. Cottingham, K. Tsu and J.-M. Richard, 
\NPB\ {\bf 179}, 541 (1981).}
%
\bibitem{Brown79}{G.E. Brown and M. Rho, {\em Phys. Lett.} {\bf 82B},
177 (1979); {\bf 84B}, 383 (1979); {\em Comments Nucl. Part. Phys.}
{\bf 18}, 1 (1988).}
%
\bibitem{Thomas84}{A.W. Thomas, in {\em Adv. Nucl. Phys.}, 
eds. J. Negele and E. Vogt, vol. {\bf 13} (1984).}
%
\bibitem{Weber86}{M. Weyrauch and H.J. Weber, \PLB\ {\bf 171}, 13
(1986); H.J. Weber and H.T. Williams, \PLB\ {\bf 205}, 118 (1988);
see, also F. Fern\'andez and E. Oset, \NPA\ {\bf 455}, 720 (1986);
I.T. Obukhovski and A.M. Kusainov, \PLB\ {\bf 328}, 142 (1990);
 A. Buchmann, E. Hern\'andez and K. Yazaki, \NPA\ {\bf 569}, 661
(1994).}
%
\bibitem{Glozman96a}{L.Ya. Glozman and D.O. Riska, {\em Phys. Rep.}
{\bf 268}, 263 (1996).}
%
\bibitem{Tornqvist91}{N. T{\"o}rnqvist, \PRL\ {\bf 67}, 556 (1991); 
\ZPC\ {\bf 61}, 525 (1994).}
%
\bibitem{Manohar93}{A.V. Manohar and M.B. Wise, 
\NPB\ {\bf 399}, 17 (1993).}
%
\bibitem{Ericson93}{T.E.O. Ericson and G. Karl, 
\PLB\ {\bf 309}, 426 (1993).}
%
\bibitem{Ader82}{J.-P. Ader, J.-M. Richard and P. Taxil,
\PRD\ {\bf 25}, 2370 (1982).}
%
\bibitem{Zouzou86}{S.~Zouzou, B. Silvestre-Brac, C. Gignoux and J.-M.
Richard,
\ZPC\ {\bf 30}, 457 (1986); L. Heller and J.A. Tjon, \PRD\ {\bf 32},
755 (1985); {\bf 35}, 969 (1987); J. Carlson, L. Heller and J.A.
Tjon, \PRD\ {\bf 37}, 744 (1987); H.J. Lipkin, \PLB\ {\bf 172}, 242
(1987).}
%
\bibitem{Silvestre93}{B. Silvestre-Brac and C. Semay, \ZPC\ {\bf 59},
457 (1993);  {\bf 61}, 271 (1994).}
%
\bibitem{Brink94}{D.M. Brink and Fl. Stancu, \PRD\ {\bf 49}, 4665 (1994).}
%
\bibitem{Pepin96}{S. Pepin, Fl. Stancu, M. Genovese and J.-M.
Richard, in preparation.}
%
\end{thebibliography}
\end{document}